\documentclass[10pt, twocolumn, notitlepage, superscriptaddress, prl, longbibliography]{revtex4-2}

\usepackage{hyperref} 
\hypersetup
{
	colorlinks=true,       
	linkcolor=blue,        
	citecolor=blue,        
	urlcolor=blue           
} 

\usepackage{amsmath} 
\usepackage{amssymb}
\usepackage{multirow}   
\usepackage{verbatim}
\usepackage{bm}
\usepackage{graphicx} 
\usepackage{dcolumn} 
\usepackage{kotex}
\usepackage{subfigure} 
\usepackage{makecell}

\usepackage{here}
\usepackage{braket}         
\usepackage{lettrine}
\usepackage{color}
\usepackage{times}
\usepackage{hhline}
\usepackage{tabularx}
\usepackage{calc}
\usepackage{soul}
\usepackage{xcolor}
\usepackage{orcidlink}
\usepackage{comment}

\begin{document}
\title{Dyakonov-Perel-like Orbital and Spin Relaxations in Centrosymmetric Systems}

\author{Jeonghun Sohn$^{\dagger}$\,\orcidlink{0009-0001-6975-368X}}
\affiliation{Department of Physics, Pohang University of Science and Technology, Pohang 37673, Korea \looseness=-1}

\author{Jongjun M. Lee$^{\dagger}$\,\orcidlink{0000-0002-9786-1901}}
\affiliation{Department of Physics, Pohang University of Science and Technology, Pohang 37673, Korea \looseness=-1}

\author{Hyun-Woo Lee\,\orcidlink{0000-0002-1648-8093}}
\email{hwl@postech.ac.kr}
\affiliation{Department of Physics, Pohang University of Science and Technology, Pohang 37673, Korea \looseness=-1}

\begin{abstract}
The Dyakonov-Perel (DP) mechanism of spin relaxation has long been considered irrelevant in centrosymmetric systems since it
was developed originally for non-centrosymmetric ones. We investigate whether this conventional understanding extends to the realm of orbital relaxation, which has recently attracted significant attention. Surprisingly, we find that orbital relaxation in centrosymmetric systems exhibits the DP-like behavior in the weak scattering regime.
Moreover, 
the DP-like orbital relaxation can make the spin relaxation in centrosymmetric systems DP-like through the spin-orbit coupling. We also find that the DP-like orbital and spin relaxations are anisotropic even in materials with high crystal symmetry (such as face-centered cubic structure) and may depend on the orbital and spin nature of electron wavefunctions.

\end{abstract}

\maketitle

{\it Introduction.---}
Electron orbital angular momentum (OAM) dynamics in solids is getting increased attention recently~\cite{Go2021}.
Theories~\cite{Bernevig2005,Tanaka2008,Go2018} predict the orbital Hall effect even in materials with strong orbital quenching and weak spin-orbit coupling (SOC). Recent experiments~\cite{Lee2021, Lee2021NCOMM, Choi2023, Hayashi2023, Sala2023, Lyalin2023, Seifert2023} verified various theoretical predictions~\cite{Jo2018,Salemi2022,Go2020,Go2020SOC} on the OAM. Experiments also revealed surprising stability of the OAM dynamics against disorder~\cite{Choi2023,Hayashi2023,Sala2023, Lyalin2023,Seifert2023,Idrobo2024}, necessitating studies on disorder effects. Disorder effects on the orbital current generation are examined theoretically~\cite{Bernevig2005,Tanaka2008,Tang2024,liu2024dominance,canonico2024realspace}.
Another urgent issue is to understand disorder effects on orbital relaxation~\cite{Rang2024}.

The first primary objective of our paper is to theoretically understand the scattering effect on orbital relaxation in centrosymmetric systems.
In the case of spin, its relaxation is typically attributed to two  mechanisms~\cite{Zutic2004,Wu2010,Dyakonov1972,DYAKONOV1984,Fabian2007,Elliott1954,Yafet1963}. 
One mechanism is the Elliott-Yafet (EY) mechanism~\cite{Elliott1954,Yafet1963}, where scattering events alter the spin direction, leading to the shorter spin relaxation time $\tau_{\rm s}$ as the momentum scattering time $\tau_{\rm m}$ becomes shorter ($\tau_{\rm s} \propto \tau_{\rm m}$). 
The other is the Dyakonov-Perel (DP) mechanism~\cite{Dyakonov1972,DYAKONOV1984,Fabian2007}, where 
the momentum-dependent effective magnetic field (spin texture) induces the spin precession and dephasing, which is intervened by scattering. Thus, $\tau_{\rm s}$ becomes {\it longer} as $\tau_{\rm m}$ becomes shorter ($\tau_{\rm s} \propto 1/\tau_{\rm m}$).
The momentum-dependent effective field arises from the spin-momentum coupling such as Rashba and Dresselhaus couplings, which are forbidden in centrosymmetric systems. Thus, the spin relaxation in centrosymmetric systems is commonly attributed to the EY mechanism~\cite{Niimi2013,Sagasta2016,Nguyen2016,Nguyen2014}.

Our analytic and numerical calculations show that the orbital relaxation time $\tau_{\rm o}$ in centrosymmetric systems exhibits the DP-like relaxation ($\tau_{\rm o}$ increases with decreasing $\tau_{\rm m}$) for sufficiently long $\tau_{\rm m}$. That is, scattering helps electron maintain their OAM for a longer time. This is the first main result of our paper. The DP-like behavior in centrosymmetric systems contradicts the common expectation that disorder scattering accelerates orbital relaxation and is also in clear contrast to the aforementioned common wisdom for spin relaxation.
When $\tau_{\rm m}$ is sufficiently short, we find that $\tau_{\rm o}$ exhibits the EY-like relaxation ($\tau_{\rm o}$ increases with increasing $\tau_{\rm m}$).

Our second primary objective is to understand spin relaxation when the SOC exists.
When the spin ${\mathbf S}$ is
coupled to the OAM ${\mathbf L}$, the spin relaxation may be
influenced by the orbital relaxation, raising the possibility that even spin may exhibit the DP-like relaxation behavior when $\tau_{\rm m}$ is sufficiently long. Our analytic and numerical calculations confirm this possibility. This is the second main result of our paper and provides a possible explanation for the DP-like spin relaxation in centrosymmetric metal Pt~\cite{Freeman2018}.  
Our result differs from a model calculation~\cite{Boross2013}, which predicts the DP spin relaxation in centrosymmetric systems when $\tau_{\rm m}$ is sufficiently short.

\begin{figure}[t]
    \centering
    \includegraphics[width=8.5cm]{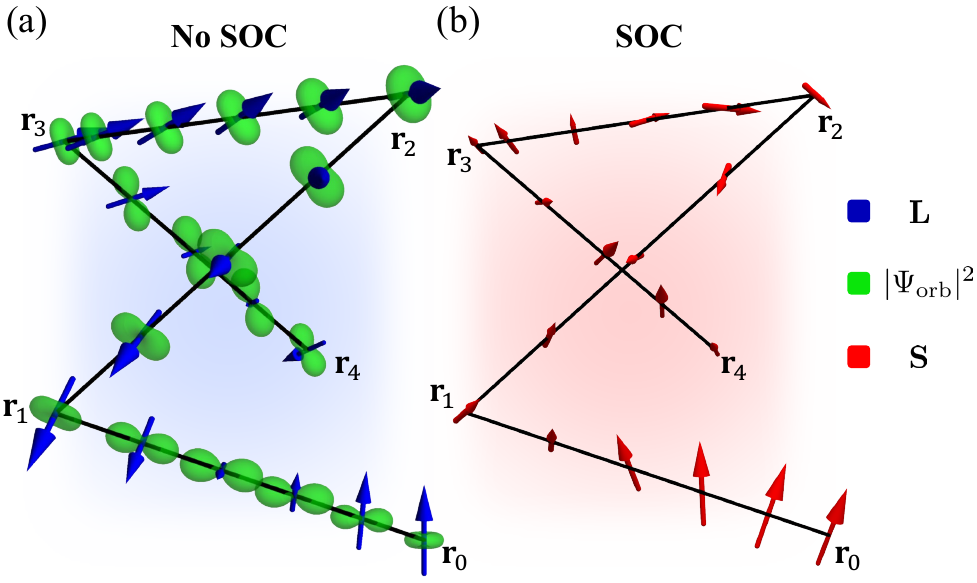}
    \caption{Schematic illustration of the DP-like relaxation. An electron is scattered at positions $\textbf{r}_1$, $\textbf{r}_2$, etc, and forms a zigzag trajectory. (a) Along the trajectory, the electron's orbital density (green) and OAM (blue arrows) evolve, which results in the DP-like orbital relaxation even in centrosymmetric systems. Without the SOC, the orbital dynamics does not affect the spin dynamics. (b) When the SOC exists
    the spin (red arrows) dynamics is affected by the orbital dynamics. Thus the OAM in (a) can cause the spin to precess, leading to the DP-like spin relaxation. 
    } 
    \label{FIG1}
\end{figure}

{\it Schematic illustration.---}
In centrosymmetric systems without the SOC, energy eigenstates have vanishing OAM expectation values. However, the orbital nature of the eigenstates varies with the momentum ${\bf k}$. 
This orbital texture can induce the DP-like OAM relaxation as demonstrated below.
For concreteness of illustration, we take a $p$ orbital centrosymmetric system with three energy bands ($n=1,2,3$), whose eigenstates have real orbital characters $|\chi_{n{\bf k}}\rangle=\hat{\bf m}_{n{\bf k}}\cdot\hat{\bf x}|p_x\rangle+\hat{\bf m}_{n{\bf k}}\cdot\hat{\bf y}|p_y\rangle+\hat{\bf m}_{n{\bf k}}\cdot\hat{\bf z}|p_z\rangle$ with $\hat{\bf m}_{n{\bf k}} \cdot \hat{\bf m}_{n'{\bf k}}=\delta_{n,n'}$. Here, the ${\bf k}$-dependence of the unit vector $\hat{\bf m}_{n{\bf k}}$ describes the orbital texture.
For a schematically illustrated electron trajectory in Fig.~\ref{FIG1}(a), we focus on a ballistic segment from ${\bf r}_a$ ($a=$0, 1, 2, $\cdots$) to ${\bf r}_{a+1}$, where the electron wavefunction for a given energy $E$ may be expressed as
\begin{equation}
    \psi_a({\bf r})=\sum_n c_{n,a} \exp\left(i{\bf k}_{n,a}\cdot  {\bf r}\right)|\chi_{n{\bf k}_{n,a}}\rangle.
\end{equation}
The $n$-dependence of ${\bf k}_{n,a}$ in the exponential phase factor arises since each band component has a different wavevector for the given $E$. 
Then, the local OAM expectation value for $\psi_a({\bf r})$ reduces to $\langle {\bf L} \rangle_{\bf r}\equiv \psi_a^\dagger ({\bf r}) {\bf L} \psi_a({\bf r})$ at ${\bf r}$, where ${\bf L}$ is the OAM operator. Although the OAM expectation value for each eigenstate $e^{i{\bf k}\cdot {\bf r}}|\chi_{n{\bf k}}\rangle$ vanishes for all ${\bf r}$, we emphasize that $\langle {\bf L} \rangle_{\bf r}$ is finite in general and varies as the electron propagates from ${\bf r}_a$ to ${\bf r}_{a+1}$. For simplicity, we take $\hat{\bf m}_{n=1,{\bf k}_{n=1,a}}=\hat{\bf x}$, $\hat{\bf m}_{n=2,{\bf k}_{n=2,a}}=\hat{\bf y}$, $\hat{\bf m}_{n=3,{\bf k}_{n=3,a}}=\hat{\bf z}$. Then, the three components $\langle L_x \rangle_{\bf r}$, $\langle L_y \rangle_{\bf r}$, and $\langle L_z \rangle_{\bf r}$ oscillate with the wavevectors ${\bf k}_{n=2,a}-{\bf k}_{n=3,a}$, ${\bf k}_{n=3,a}-{\bf k}_{n=1,a}$, and ${\bf k}_{n=1,a}-{\bf k}_{n=2,a}$, respectively. This illustrates that the OAM can precess intrinsically due to the orbital texture, which is allowed even in centrosymmetric systems~\cite{Han2023}.

To understand the scattering effect on the intrinsic OAM dynamics,
we note that the $\langle{\mathbf{L}}\rangle_{\bf r}$ evolution pattern depends crucially on the eigenstate orbital characters, which vary with the propagation direction due to the orbital texture. This implies that $\langle{\mathbf{L}}\rangle_{\bf r}$ evolution pattern in a ballistic segment between two consecutive scatterings at $\mathbf{r}_{j}$ and $\mathbf{r}_{j+1}$ differs from the corresponding patterns in different ballistic segments. Therefore, each scattering alters the $\langle{\mathbf{L}}\rangle_{\bf r}$ evolution pattern [Fig.~\ref{FIG1}(a)],
which resembles the $\langle \mathbf{S}\rangle_{\bf r}$ evolution pattern change by scattering in noncentrosymmetric systems. This implies that the DP-like OAM relaxation may occur in centrosymmetric systems. 

Next, we turn on the SOC. Then, the local spin expectation value $\langle{\mathbf{S}}\rangle_{\bf r}$ should also evolve intrinsically since ${\bf S}$ is coupled to ${\bf L}$ and $\langle{\mathbf{L}}\rangle_{\bf r}$ evolves intrinsically. Moreover, $\langle{\mathbf{S}}\rangle_{\bf r}$ evolution pattern should change upon scattering [see the evolution of the red arrows in Fig.~\ref{FIG1}(b)] just like $\langle{\mathbf{L}}\rangle_{\bf r}$ does. This illustrates schematically that the DP-like spin relaxation may be possible in centrosymmetric systems when the SOC exists.

{\it Kinetic theory approach.---} 
To gain a more rigorous assessment of the orbital relaxation, we apply the semiclassical kinetic theory to a centrosymmetric $p$-orbital model system described by the Luttinger Hamiltonian~\cite{Luttinger1956}, which was used previously to demonstrate the intrinsic orbital Hall effect in hole-doped silicon~\cite{Bernevig2005}. Its Hamiltonian density is given by,
\begin{equation}
\begin{aligned}
\mathcal{H}^{\rm o}_{\bf k}=c_0 {\bf k}^2+c_1 \sum\nolimits_{j}k_j^2 L_j^2
    +c_{2} \sum\nolimits_{i\neq j}k_{i}k_{j}\{ {L}_{i},{L}_{j} \},
\end{aligned}
\label{Ham_L1}
\end{equation}
where $k_i$ and $L_{i}$ denote components of ${\bf k}$ and ${\bf L}$, respectively, and 
$i,j=x,y,z$. 
Note that there is no SOC in $\mathcal{H}^{\rm o}_{\mathbf{k}}$. In the first term, $\mathbf{k}$ does not couple to ${\mathbf{L}}$, whereas in the second and the third terms, $\mathbf{k}$ couples to ${\mathbf{L}}$, generating the $\mathbf{k}$-dependent variation of the real orbital character $|\chi_{n{\bf k}}\rangle$ (orbital texture~\cite{Go2018}). These forms of orbital-momentum coupling do not break the inversion symmetry nor the time-reversal symmetry~\cite{Han2022,Luttinger1956}.

Our semiclassical kinetic theory builds upon the original work of Dyakonov and Perel~\cite{Dyakonov1972,DYAKONOV1984,Fabian2007}, who solved the quantum Boltzmann equation and demonstrated the DP spin relaxation due to the spin texture in noncentrosymmetric systems. 
To assess the effect of the orbital texture on orbital relaxation in centrosymmetric systems,
we treat the orbital-momentum coupling terms in $\mathcal{H}^{\rm o}_{\mathbf{k}}$ as perturbations and the first term in $\mathcal{H}^{\rm o}_{\mathbf{k}}$ as an unperturbed Hamiltonian. In the interaction picture, the equation of motion for the density matrix $\rho_{\bf k}$ then reads,
\begin{equation}
    \frac{\partial {\rho}_{\mathbf{k}}}{\partial t} = \frac{1}{i\hbar}[\mathcal{H}^{\text{om}}_{\mathbf{k}},{\rho}_{\mathbf{k}}] + \Big(\frac{\partial {\rho}_{\mathbf{k}}}{\partial t}\Big)_{\text{coll}},
\label{Eq_kinetic_1}
\end{equation}
where 
$\mathcal{H}^{\text{om}}_{\mathbf{k}}$ denotes the orbital-momentum coupling terms in Eq.~(\ref{Ham_L1}). The commutator on the right-hand side of Eq.~(\ref{Eq_kinetic_1}) induces the intrinsic time evolution of the orbital. The collision integral by impurities [last term in Eq.~(\ref{Eq_kinetic_1})] is assumed to be elastic and isotropic. Furthermore, we assume scattering-induced orbital relaxation is weak, which can be justified in the large $\tau_{\rm m}$ regime, where the EY-like orbital relaxation is negligible.
We investigate the evolution of the OAM, $\text{Tr}[{\rho}_{\mathbf{k}}{\mathbf{L}}]$, by solving the corresponding equation and taking its angular average over the direction $\mathbf{k}$. As a result, we find the orbital relaxation time $\tau_{\rm o}$ given by,
\begin{equation}
    \frac{1}{\tau_{\rm o}} = f_{\rm o} \tau_{\rm m},
\label{Eq_Orbital_DP_inv}
\end{equation}
where $f_{\rm o}= (2k^{4}_{\rm F}/15\hbar^{2} \gamma_{\rm o})(2c^{2}_{1}+3c^{2}_{2})$, 
$k_{\rm F}$ is the momentum at the Fermi surface,
and $\gamma_{\rm o}$ is a dimensionless constant of order one that does not depend on $\tau_{\rm m}$~\cite{suppl_ref,DYAKONOV1984,Fabian2007}.
Note that $1/\tau_{\rm o}$ is proportional to $\tau_{\rm m}$; that is, the orbital relaxation follows the DP behavior.

Next, we introduce the SOC ${\bf S}\cdot{\bf L}$ to $\mathcal{H}^{\rm o}_{\bf k}$ and examine the spin relaxation in a centrosymmetric $p$-orbital system described by
\begin{equation}
    \mathcal{H}^{\rm os}_{\mathbf{k}}
    =c_0 {\bf k}^2
    +c_1 \left( {\bf k} \cdot {\bf L} \right)^2
    +g_{\text so}{\bf S}\cdot {\bf L},
\label{Eq_H_1}
\end{equation}
where the third term is the SOC.
The isotropic limit ($c_1 = 2c_2$) is assumed for simplicity.
With $\mathcal{H}^{\rm os}_{\mathbf{k}}$, we perform a similar semi-classical calculation of the quantum Boltzmann equation. We treat the SOC as a perturbation, noting that the orbital-momentum coupling has a typical energy scale comparable with the electron hopping and is stronger than the SOC. 
We investigate the evolution of the spin, $\text{Tr}[{\rho}_{\mathbf{k}}{\mathbf{S}}]$, and take its angular average over the direction $\mathbf{k}$. As a result, we find that in the large $\tau_{\rm m}$ regime, the spin relaxation time $\tau_{\rm s}$ is given by
\begin{equation}
    \frac{1}{\tau_{\rm s}} = f_{\rm s} \tau_{\rm m},
\label{Eq_DP_spin_inv}
\end{equation}
where $f_{\rm s} = (16/3 \hbar^{2}\gamma_{\rm s}) g^{2}_{\text{so}}$ and $\gamma_{\rm s}$ is a dimensionless constant of order one that does not depend on $\tau_{\rm m}$~\cite{suppl_ref,DYAKONOV1984,Fabian2007}.
Note that $1/\tau_{\rm s}$ is proportional to $\tau_{\rm m}$; that is, the spin relaxation follows the DP behavior even though the spin-momentum coupling is forbidden in this centrosymmetric system. This DP behavior 
is realized by the combination of the orbital texture and the SOC, as evidenced by $f_{\rm s}$ being proportional to $g^{2}_{\text{so}}$. Our results demonstrate that the orbital texture can induce the DP-like relaxation of both spin and orbital in centrosymmetric nonmagnetic systems.

\begin{figure}[t]
    \centering
    \includegraphics[width=\linewidth]{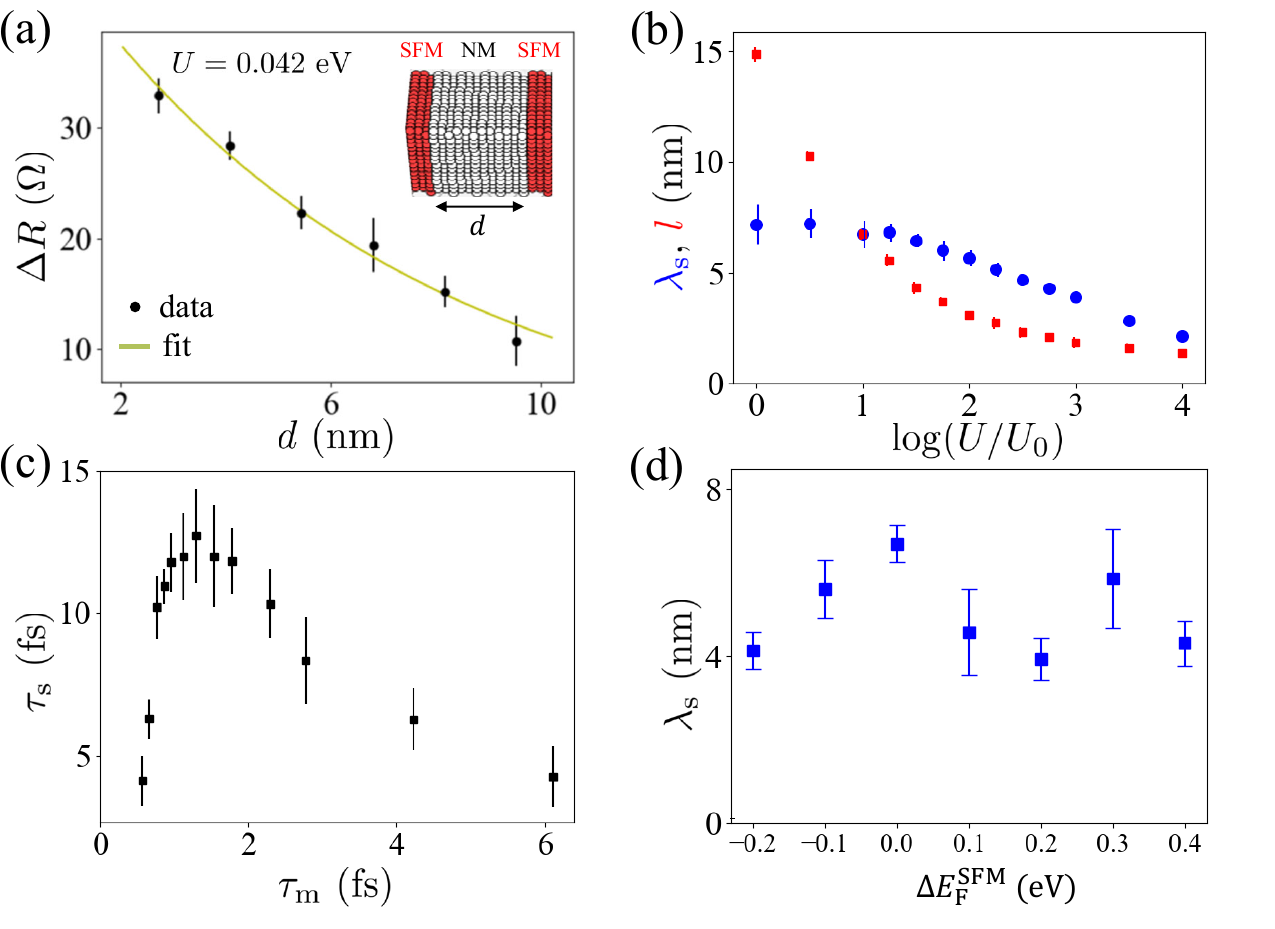}
    \caption{Spin relaxation in NM. (a) Magnetoresistance $\Delta R$ as a function of the NM thickness $d$ with random static disorder potential with magnitude $U=0.042$ eV. Points denote $\Delta R$
    averaged over disorder configurations. The vertical bars denote the standard deviation of $\Delta R$. The curve denotes the exponential fitting of the points. Inset: Schematic of the spin valve structure. (b) Spin relaxation length $\lambda_{\rm s}$ (blue) and mean free path $l$ (red) as a function of $U$. 
    (c) Spin relaxation time $\tau_{\rm s}$ as a function of momentum scattering time $\tau_{\rm m}$.
    (d) $\lambda_{\rm s}$'s dependence on the onsite energy shift in SFMs.
    }
    \label{Spin_Relax}
\end{figure}

\begin{figure}[t!]
    \centering
    \includegraphics[width=\linewidth]{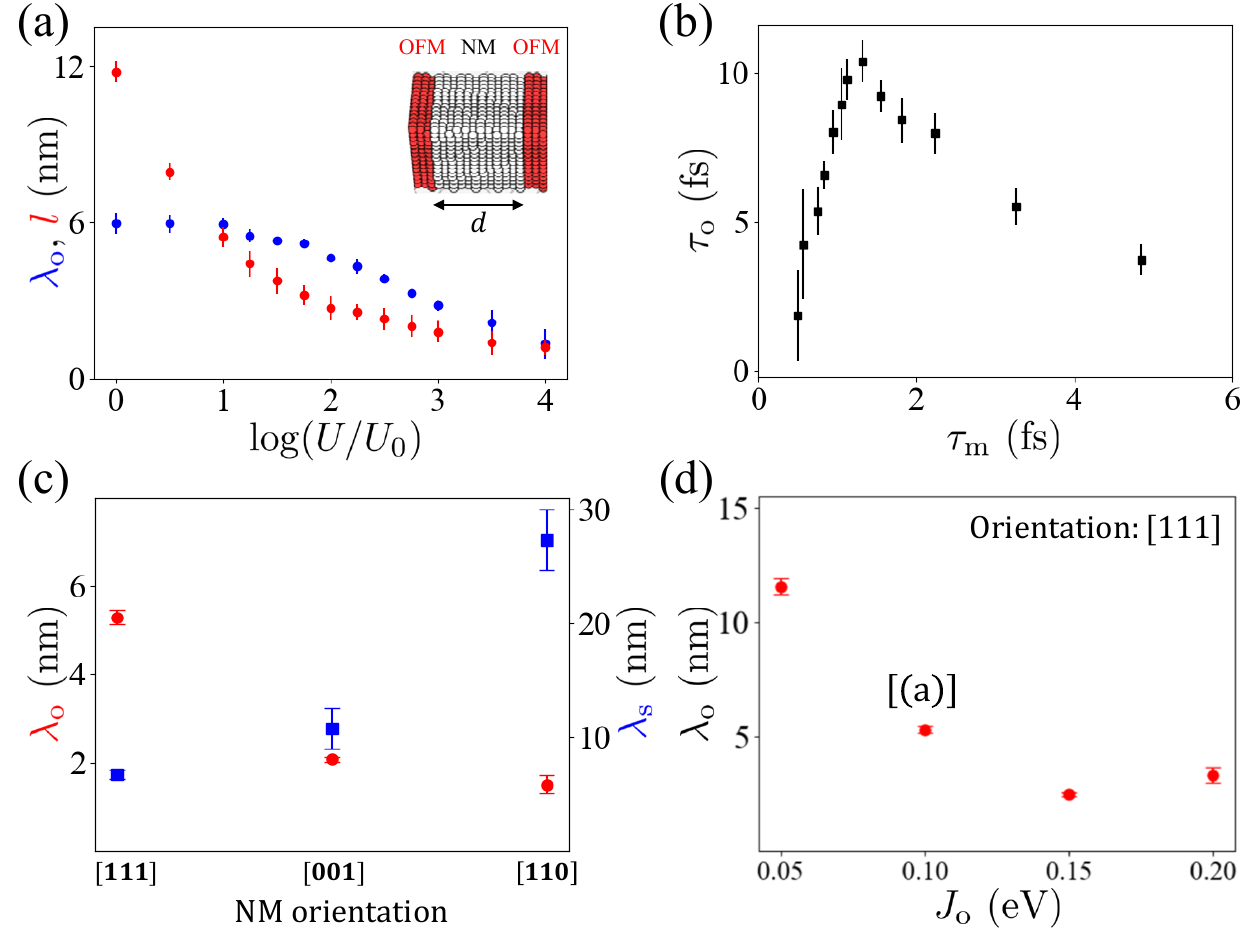}
    \caption{Orbital relaxation in NM. (a) Orbital relaxation length $\lambda_{\rm o}$ (blue) and mean free path $l$ (red) as a function of $U$. Inset: orbital valve structure. 
    (b) Orbital relaxation time $\tau_{\rm o}$ as a function of momentum scattering time $\tau_{\rm m}$. 
    (c) $\lambda_{\rm o}$ and $\lambda_{\rm s}$ for several crystal directions of NM.
    (d) $\lambda_{\rm o}$ for different values of orbital exchange coupling parameter $J_{\rm o}$. $U=0.042$ eV in (c) and (d).} 
    \label{Orbital_relax}
\end{figure}

{\it Numerical calculation.---}
We examine the spin and orbital relaxations in centrosymmetric nonmagnetic metals (NM) numerically by using the tight-binding description. 
We calculate the spin-dependent or orbital-dependent {\it charge} conductances for a spin value [Fig.~\ref{Spin_Relax}(a) inset] or an orbital valve [Fig.~\ref{Orbital_relax}(a) inset], where the NM of interest is sandwiched between two spin ferromagnets (SFMs) or two orbital ferromagnets (OFMs)~\cite{Tschirhart2021,Shindou2001}, which act as spin or OAM generators and detectors. We calculate the charge resistance $R$ between the two ferromagnets (FM) for the two configurations, one with the two FMs magnetized parallel to each other ($R_{\rm P}$) and the other with the two FMs magnetized antiparallel ($R_{\rm AP}$). The resistance difference $\Delta R=R_{\rm AP}-R_{\rm P}$ amounts to the out-of-plane giant magnetoresistance, which decays as the distance $d$ between the two FMs increases. We extract the spin and orbital relaxation lengths of the NM from the $d$-dependence of $\Delta R$, which changes with the disorder strength in the NM. 
This scheme allows one to focus on the scattering effect on longitudinal spin or orbital current relaxation in the NM and to decouple it from the scattering effect on spin~\cite{Murakami2004,Tanaka2008} or orbital~\cite{Bernevig2005,Tanaka2008,Tang2024,liu2024dominance,canonico2024realspace} Hall current generation in the NM since the latter effect does not affect $\Delta R$.
The relaxation lengths are later converted to the spin and orbital relaxation times.   

Since the modeling of the SFMs is more straightforward than the OFMs, we examine the spin relaxation in a spin valve structure first [Fig.~\ref{Spin_Relax}(a) inset]. For both the left and right SFMs in the spin valve, we adopt artificial FMs~\cite{suppl_ref} emulating permalloy Ni$_{81}$Fe$_{19}$, since permalloy is commonly used to investigate spin relaxation~\cite{Freeman2018,Sagasta2016,Kimura2008}. For the NM in the spin valve, we adopt the Slater-Koster tight-binding parameters of Pt. Additionally, we introduce static on-site disorder potential with a Gaussian random distribution (mean = 0, standard deviation = $U$) to induce scattering in the NM. No disorder is introduced in the SFMs.
All three layers in the spin valve
possess the face-centered-cubic structure stacked along the [111] direction. The magnetization directions of the SFMs are perpendicular to the interfaces. 
%
Using the \textsc{Kwant} package~\cite{Groth2014,Amestoy2006}, we evaluate the average (symbols) and the standard deviation (vertical bars) [Fig.~\ref{Spin_Relax}(a)] of $\Delta R$
over the on-site disorder potential realizations, for given values of $U$ and $d$. 
The $d$-dependence of the average is fitted by the curve (solid line),
\begin{align} \label{fit}
    \Delta R_{\rm ave} = \Delta R_{0}e^{-d/\lambda_{\rm s}}
\end{align}
where $\lambda_{\rm s}$ is the spin relaxation length and $\Delta R_{0}$ is a $d$-independent constant. 

We then repeat the process in Fig.~\ref{Spin_Relax}(a) for various values of $U$ to analyze how $\lambda_{\rm s}$ varies with $U$ [blue symbols in Fig.~\ref{Spin_Relax}(b)]. The mean free path ($l$) is also evaluated as a function $U$ [red symbols in Fig.~\ref{Spin_Relax}(b)]. To determine $l$, we calculate the disorder-averaged resistivity of the NM~\cite{suppl_ref} and convert it into $l$ and $\tau_\mathrm{m}$. 
For $\log (U/U_0)\lesssim 2$ ($U_0=0.03$ eV), we find that $l$ decays rapidly as $U$ increases whereas $\lambda_{\rm s}$ decreases only mildly with $U$. This difference in the $U$-dependences of $l$ and $\lambda_{\rm s}$ implies that the spin relaxation time $\tau_{\rm s}$ {\it increases} as $U$ increases. This confirms that even in centrosymmetric systems, the DP-like spin relaxation is possible in the weak scattering regime. 
%
To quantify $\tau_{\rm s}$, we utilize the relation $\tau_{\rm s} = 3\lambda_{\rm s}^{2} v^{-2}_{\rm F} \tau_\mathrm{m}^{-1}$~\cite{Bass2007}, where $v_{\rm F}$ is the Fermi velocity. This relation is derived in the diffusive regime ($\lambda_{\rm s} \gg l$), and thus not strictly valid when $\lambda_{\rm s}$ is comparable to $l$ [Fig.~\ref{Spin_Relax}(b)]. Nevertheless, we use the relation since the diffusion picture is commonly used in the spin transport analysis 
even when $\lambda_{\rm s}\sim l$~\cite{Yang1994,Bass1994,Fert1995,Freeman2018}. 
%
Figure~\ref{Spin_Relax}(c) shows the resulting $\tau_{\rm s}$ plotted as a function of $\tau_{\rm m}$.
In the weak scattering regime, 
($\tau_{\rm m}\geq 1.5$ fs), $\tau_{\rm s}$ increases with decreasing $\tau_\mathrm{\rm m}$, thus exhibiting the DP-like spin relaxation.
On the other hand, in the strong scattering regime ($\tau_{\rm m}\leq 1.5$ fs), $\tau_{\rm s}$ increases with increasing $\tau_{\rm m}$, thus exhibiting the EY-like spin relaxation.

Next, we examine orbital relaxation numerically via an orbital valve structure [Fig.~\ref{Orbital_relax}(a) inset] with an NM sandwiched between two OFMs.
All three layers in the orbital valve possess the face-centered-cubic structure stacked along the [111] direction. For the NM, we turn off the SOC in the tight-binding parameters of Pt. For the OFMs, we adopt a simple modeling of artificial OFMs instead of real OFMs~\cite{Tschirhart2021,Shindou2001}; we introduce the orbital exchange coupling $J_{\rm o}\hat{\bf M}\cdot {\bf L}$ to the tight-binding parameters of the NM. Here the unit vector $\hat{\bf M}$ denotes the magnetization direction of an OFM and $J_{\rm o}$ is the strength of the orbital exchange coupling. Due to the coupling, eigenstates of the OFMs have finite OAM expectation values. The reversal of $\hat{\bf M}$ does not modify the band structure of the OFMs but reverses the OAM expectation values of eigenstates. The SOC is absent everywhere in the orbital valve so that we can focus on pure orbital relaxation dynamics, decoupled from spin dynamics.
The rest of the analysis procedure is the same as the spin relaxation analysis procedure.
%
Figure~\ref{Orbital_relax}(a) compares the resulting orbital relaxation length $\lambda_{\rm o}$ and $l$ as a function of $U$. Their $U$-dependences are similar to those in Fig.~\ref{Spin_Relax}(b). This confirms that orbital relaxation in centrosymmetric systems can exhibit the DP-like relaxation in the weak scattering regime.
The relation $\tau_{\rm o} = 3\lambda_{\rm o}^{2} v^{-2}_{\rm F} \tau_\mathrm{m}^{-1}$ is used to convert $\lambda_{\rm o}$ to the orbital relaxation time $\tau_{\rm o}$. The resulting $\tau_{\rm o}$ is shown as a function of $\tau_{\rm m}$ in Fig.~\ref{Orbital_relax}(b), where the DP-like and EY-like orbital relaxations are evident in the weak and strong scattering regimes, respectively. 

{\it Discussions.---} 
A few experiments~\cite{Kimura2008,Mihajlovic2010,Ryu2016,Freeman2018} reported that the spin relaxation in centrosymmetric metallic systems exhibits the DP-like behavior in the low-temperature regime (low-resistivity regime). A recent calculation~\cite{Belashchenko2023} also reported the DP-like spin relaxation in Pt. Since the DP spin relaxation has been believed to be forbidden in centrosymmetric systems, some of them~\cite{Mihajlovic2010,Ryu2016,Belashchenko2023} attributed the behavior to the interfaces of the centrosymmetric materials, where the inversion symmetry is broken and the spin-momentum coupling may be present. On the other hand, the experiment on Pt~\cite{Freeman2018} argued that the DP spin relaxation by the interfaces is not relevant in their measurement geometry of the out-of-plane magnetoresistance [same as our spin valve geometry in Fig.~\ref{Spin_Relax}(a) inset] and the observed DP-like behavior should originate from centrosymmetric bulk Pt. We argue that our work provides a possible explanation for the previously reported DP-like behaviors in centrosymmetric systems. To separate the interfacial and bulk contributions to the DP-like relaxation in our calculation, we turn off the SOC ${\bf S}\cdot{\bf L}$ in the bulk of the NM and retain the SOC only near the SFM/NM and the NM/SFM interfaces. A significantly longer spin relaxation length is obtained in this case~\cite{suppl_ref}, implying that the DP-like spin relaxation in Fig.~\ref{Spin_Relax} arises mainly from the bulk NM, consistent with the argument of Ref.~\cite{Freeman2018}. For the orbital relaxation, on the other hand, we are not aware of any experiments that systematically examine the scattering effect. We argue that such experiments are highly needed to clarify the nature of the orbital relaxation. 
 
Our numerical calculations reveal a couple of properties that go beyond the simple model Hamiltonians $\mathcal{H}^{\rm o}_{\bf k}$ [Eq.~(\ref{Ham_L1})] and $\mathcal{H}^{\rm os}_{\bf k}$ [Eq.~(\ref{Eq_H_1})], since the Hamiltonians for the numerical calculations are more complicated.
The first such property is the anisotropy of relaxation in the weak scattering regime. The anisotropy of the EY-type spin relaxation was reported before~\cite{Mavropoulos2012,Zimmermann2016}. To examine the relaxation anisotropy, we change the stacking direction of the spin and orbital valves from [111] to [001], and [110], and examine the out-of-plane magnetoresistance. The magnetization direction is maintained to be parallel or antiparallel to the stacking direction.
We find that $\lambda_{\rm s}$ ($\lambda_{\rm o}$) in the weak scattering regime ($U=0.042$ eV) varies factor 4.1 (3.5) with the stacking direction [Fig.~\ref{Orbital_relax}(c)]. 
We attribute the relaxation anisotropy to the inequality $\lambda_{\rm s}, \lambda_{\rm o} \lesssim l$
[Figs.~\ref{Spin_Relax}(b) and \ref{Orbital_relax}(a)], frequently encountered situations in experiments~\cite{Nguyen2016,Rojas-Sanchez2014,Kurt2002,Lyalin2023,Lee2021}. In such situations, electron properties are not averaged over the entire Fermi surface but instead influenced strongly by relatively narrow regions of the Fermi surface that vary with the electron propagation direction.
Interestingly, the [111] ([110]) direction has the smallest (largest) $\lambda_{\rm s}$ but the largest (smallest) $\lambda_{\rm o}$.
We attribute this behavior to the direction-dependent orbital environment change.
Along the [111] direction, the $C_3$ rotational symmetry safeguards the two-fold orbital degeneracy of $d_{xy}$ and $d_{x^2-y^2}$ character states~\cite{Fu2011} 
($z$ along [111]), facilitating the transport of $L_z=\pm 2\hbar$ orbital ($d_{x^2-y^2}\pm i d_{xy}$) and resulting in the longest $\lambda_{\rm o}$. 
The importance of the degeneracy for $\lambda_{\rm o}$ was reported for NM~\cite{Urazhdin2023} and FM~\cite{Go2023}.
In such a situation with orbital degeneracy, the SOC becomes more effective and can modify the eigenstates significantly (thereby suppressing $\lambda_{\rm s}$).
Exactly opposites apply to the [110] direction, along which a large crystal field splitting occurs. In such a situation with the lifted orbital degeneracy, the SOC becomes ineffective (thereby enhancing $\lambda_{\rm s}$).
The relaxation anisotropy also raises a possibility that for a given multilayer structure with a well-defined stacking direction, its out-of-plane transport~\cite{Kurt2002,Freeman2018} and lateral transport~\cite{Kimura2008,Valenzuela2006} may have different relaxation length scales~\cite{Bass2007}. A recent calculation on Pt~\cite{Belashchenko2023} reported that the spin relaxation lengths along the transverse and longitudinal directions are different from each other, although its relation with our result [Fig.~\ref{Orbital_relax}(c)] is unclear.  
Further study is needed to clarify the relaxation anisotropy,
which is in clear contrast to the spin (orbital) Hall conductivity that is isotropic in the face-centered cubic materials.
Also, we find that $\lambda_{\rm o}$ and $\lambda_{\rm s}$ depend not only on the material parameters of NMs but also on those of OFMs and SFMs.
Figure~\ref{Orbital_relax}(d) shows the variation of $\lambda_{\rm o}$ as the orbital exchange coupling parameter $J_{\rm o}$ changes, 
and Fig.~\ref{Spin_Relax}(d) shows the variation of $\lambda_{\rm s}$ with $\Delta E_{\rm F}^{\rm SFM}$, where $\Delta E_{\rm F}^{\rm SFM}$ denotes the onsite energy shift of all sites in SFMs. 
Both figures are obtained in the weak scattering regime ($U=0.042$ eV).
Considering that $J_{\rm o}$ and $\Delta E_{\rm F}^{\rm SFM}$ modify the wavefunction shape of the states at the Fermi energy of OFMs and SFMs, respectively, we attribute the $J_{\rm o}$- and $\Delta E_{\rm F}^{\rm SFM}$- dependences to the $\lambda_{\rm o}$- and $\lambda_{\rm s}$- dependences on the wavefunction nature of injected electrons when $\lambda_{\rm o}, \lambda_{\rm s} \lesssim l$.  
%
%
This result may provide a partial explanation as to why diverse values of $\lambda_{\rm s}$ are obtained for the same NMs in experiments~\cite{Bass2007,Rojas-Sanchez2014,Tao2018} since $\lambda_{\rm s}$ may vary with FMs in contact with NMs.

An important outstanding problem is to resolve the discrepancy between experiments reporting long $\lambda_{\rm o}$ (ranging from several nanometers~\cite{Lee2021,Idrobo2024} to tens of nanometers~\cite{Choi2023,Hayashi2023,Idrobo2024}) and calculations reporting much shorter $\lambda_{\rm o}$ (typically less than one nanometer~\cite{Salemi2021,Belashchenko2023,Urazhdin2023,Rang2024}). Although our calculation cannot resolve the discrepancy since it aims to explore qualitative behaviors of $\lambda_{\rm o}$ and $\tau_{\rm o}$ instead of their quantitative accuracy, the anisotropy [Fig.~\ref{Orbital_relax}(c)] and the FM-dependence [Fig.~\ref{Orbital_relax}(d)] of $\lambda_{\rm o}$ may be of relevance to the discrepancy. Another outstanding problem, which may be related to the above problem, is to understand orbital relaxation in polycrystalline systems.
Most experiments on orbital transport utilize polycrystalline materials~\cite{Choi2023,Hayashi2023,Sala2023,Lyalin2023,Seifert2023,Idrobo2024}, whereas polycrystallinity is ignored in all existing calculations. It was suggested~\cite{Liao2022} that orbital transport may be more efficient in polycrystalline materials than in single crystals. A recent experiment~\cite{Idrobo2024} supports this suggestion. Further studies are needed for these outstanding problems.

{\it Conclusion.---}
We demonstrated that in the weak scattering regime, both orbital and spin relaxations may exhibit DP-like behaviors even in centrosymmetric systems. The orbital texture plays a crucial role in the DP-like relaxations. 
Our results provide an explanation for a recent experiment on DP-like spin relaxation in Pt~\cite{Freeman2018} and predict the relaxation anisotropy and relaxation's dependence on the orbital and spin nature of electron wavefunctions. More experiments are needed to investigate intriguing properties of orbital and spin relaxations systematically.  

\acknowledgements
We acknowledge Jinsoo Park, Axel Hoffmann, Kyoung-Whan Kim, Insu Baek, and Hojun Lee for fruitful discussions. We acknowledge support from the Samsung Science and Technology Foundation (BA-1501-51). J.M.L. was supported by the POSCO Science Fellowship of the POSCO TJ Park Foundation. J.S. and J.M.L. contributed equally to this work$^\dagger$.

\bibliography{BibRef}

\end{document}